\documentclass[11pt,twocolumn]{article}
\usepackage[utf8]{inputenc}
\usepackage{amsmath, amssymb, amsthm}
\usepackage{graphicx}
\usepackage{hyperref}
\usepackage[margin=1in]{geometry}
\usepackage{booktabs}
\usepackage{caption}
\usepackage{float}
\usepackage{enumitem}
\setlist{nosep, leftmargin=*}
\hypersetup{colorlinks=true, citecolor=blue, urlcolor=blue}

\title{CATEKAPPA: An R Shiny Application for Design and Analysis of Consistency Tests Based on the Kappa Statistic for Categorical Responses}
\author{Zheng Gai$^{1}$, Li Xincheng$^{2}$, Jiang Wangyingjie$^{3}$, Zhao Panwei$^{4}$\\
\thanks{$^{1}$Email: \texttt{z2118778229@163.com}; $^{2}$Email: \texttt{2700753941@qq.com}; $^{3}$Email: \texttt{2312055564@qq.com}; $^{4}$Email: \texttt{1581729526@qq.com}}}
\date{\today}
\usepackage{etoolbox}
\AtBeginEnvironment{verbatim}{\small}
\usepackage{xurl}
\usepackage{listings}
\usepackage{listings}
\usepackage{xcolor}

\usepackage{tcolorbox}
\tcbuselibrary{listings,breakable}
\usepackage{verbatim}

\newtcolorbox{codebox}{
    colback=gray!5,
    colframe=gray!50,
    arc=2pt,
    boxrule=0.5pt,
    breakable,
    left=2mm,
    right=2mm,
    top=1mm,
    bottom=1mm,
    fontupper=\ttfamily\small
}
\usepackage{ragged2e}
\usepackage[expansion,protrusion]{microtype}

\RaggedRight        % 左对齐，彻底消灭大空格 + 不溢出
\begin{document}

\maketitle
\begin{abstract}
The kappa statistic is the most widely used measure of inter-rater agreement for categorical data. Despite its popularity, applied researchers often encounter two major hurdles: (i) determining the sample size required to achieve a desired level of agreement with given power, and (ii) computing appropriate kappa coefficients with proper interpretation. Existing R packages such as \texttt{irr} and \texttt{kappaSize} provide these functionalities but require programming skills and lack an integrated, user-friendly interface. We present \texttt{CATEKAPPA}, an R package that bridges this gap by combining sample size planning (via \texttt{kappaSize}) and agreement analysis (via \texttt{irr}) into a single Shiny-based web application. The package supports Cohen's kappa (2 raters), Fleiss' kappa ($\ge 3$ raters), and Light's kappa, and provides automatic interpretation using the Landis \& Koch scale. Users can either launch an interactive graphical interface or use command-line functions for scripting. The package is freely available on CRAN.
\end{abstract}

\section{Introduction}

In many scientific domains—clinical medicine, psychology, content analysis, and social sciences—the reliability of categorical ratings by multiple observers is fundamental. For example, two pathologists may classify tumor biopsies as ``benign'' or ``malignant''; several psychiatrists may diagnose a patient's mental disorder into nominal categories; or multiple coders may label social media posts as ``positive'', ``neutral'', or ``negative''. In all these settings, quantifying the extent to which raters agree beyond chance is essential for establishing measurement credibility.

Since its introduction by \cite{cohen1960coefficient}, the kappa statistic has become the standard tool for measuring inter-rater agreement for nominal scales. Extensions for multiple raters were later developed by \cite{fleiss1971measuring} and \cite{light1971measures}. Moreover, sample size planning for kappa-based hypothesis tests was studied by \cite{flack1988sample} and \cite{rotondi2012confidence}. Landis and Koch \cite{landis1977measurement} provided widely adopted benchmarks for interpreting kappa values.

Despite the availability of several R packages (e.g., \texttt{irr} \cite{irrpackage}, \texttt{kappaSize} \cite{kappaSize}), two major obstacles persist for non-programmers:
\begin{enumerate}
    \item \textbf{Sample size determination}: How many subjects are needed to test $H_0: \kappa \le \kappa_0$ versus $H_1: \kappa > \kappa_0$ with desired power? This requires specifying null/alternative kappa, category proportions, significance level, power, and number of raters.
    \item \textbf{Agreement analysis}: Given observed data, what is the estimated kappa, its confidence interval, and how should it be interpreted according to established guidelines?
\end{enumerate}
No existing package seamlessly integrates both stages with a point-and-click interface.

To fill this gap, we developed \texttt{CATEKAPPA} (Categorical Agreement Test Evaluation), an R package that provides:
\begin{itemize}
    \item A Shiny-based graphical user interface (GUI) for both sample size design and consistency analysis.
    \item Command-line functions for users who prefer scripting.
    \item Support for 2 to 5 categories and 2 to 6 raters in sample size calculations (via \texttt{kappaSize}).
    \item Three kappa types for analysis: Cohen's, Fleiss', and Light's.
    \item Automatic interpretation using the Landis \& Koch scale with color-coded output.
\end{itemize}

This paper describes the statistical methodology underlying \texttt{CATEKAPPA}, its package architecture, core functions, and illustrative usage examples. The package is publicly available on CRAN \footnote{\url{https://CRAN.R-project.org/package=catekappa}} and the source code can be found at \url{https://github.com/satellite837/catekappa}.

\section{Implemented Methodology}

\subsection{Cohen's Kappa for Two Raters}

For two raters and $k$ nominal categories, let $n_{ij}$ be the number of subjects assigned to category $i$ by rater 1 and category $j$ by rater 2, with $i,j=1,\dots,k$. Define $p_{ij}=n_{ij}/N$, where $N$ is the total number of subjects. The observed agreement proportion is $p_o = \sum_{i=1}^k p_{ii}$, and the chance-expected agreement under independence is $p_e = \sum_{i=1}^k p_{i\cdot} p_{\cdot i}$, where $p_{i\cdot}$ and $p_{\cdot i}$ are marginal proportions. Cohen's kappa is then:
\[
\kappa = \frac{p_o - p_e}{1 - p_e}.
\]
The large-sample variance of $\hat{\kappa}$ was derived by \cite{fleiss1969largesample} and is implemented in the \texttt{irr} package.

\subsection{Fleiss' Kappa for Multiple Raters}

When $m\ge 3$ raters classify $N$ subjects into $k$ categories, let $n_{ij}$ be the number of raters who assigned subject $i$ to category $j$ ($i=1,\dots,N; j=1,\dots,k$). For subject $i$, the proportion of agreement among raters is $p_i = \frac{1}{m(m-1)} \sum_{j=1}^k n_{ij}(n_{ij}-1)$. The overall observed agreement is $\bar{p} = \frac{1}{N}\sum_{i=1}^N p_i$. The chance-expected agreement is $\bar{p}_e = \sum_{j=1}^k \bigl(\frac{1}{Nm}\sum_{i=1}^N n_{ij}\bigr)^2$. Then Fleiss' kappa is:
\[
\kappa = \frac{\bar{p} - \bar{p}_e}{1 - \bar{p}_e}.
\]

\subsection{Light's Kappa}

Light's kappa \cite{light1971measures} computes all pairwise Cohen's kappa values among $m$ raters and averages them. It is particularly useful when the number of raters is small ($\ge 3$) but the researcher wishes to summarize overall agreement without assuming that all raters share identical marginal distributions.

\subsection{Sample Size Determination for Two Raters}

Following \cite{flack1988sample} and the implementation in \texttt{kappaSize}, the required sample size $N$ to test $H_0:\kappa \le \kappa_0$ versus $H_1:\kappa = \kappa_1$ with significance level $\alpha$ and power $1-\beta$ is:
\[
N \ge \left[ \frac{z_{1-\alpha}\, \tau(\kappa_0) + z_{1-\beta}\, \tau(\kappa_1)}{\kappa_1 - \kappa_0} \right]^2,
\]
where $\tau(\kappa)$ is the asymptotic standard deviation factor that depends on the marginal category proportions $\pi_{1},\dots,\pi_{k}$ and the number of raters $m$. For $k=2$ categories, the function reduces to a simpler binary case; for $k=3,4,5$, \texttt{kappaSize} implements the corresponding formulas from \cite{rotondi2012confidence}. Our wrapper \texttt{calc\_sample\_size\_kappa} automatically selects the appropriate \texttt{kappaSize} function based on the length of the \texttt{props} argument.

\subsection{Interpretation of Kappa}

We adopt the Landis and Koch \cite{landis1977measurement} benchmarks, as shown in Table~\ref{tab:landis}.

\begin{table}[H]
\centering
\caption{Landis \& Koch interpretation scale}
\label{tab:landis}
\begin{tabular}{lc}
\toprule
Kappa range & Strength of agreement \\
\midrule
$<0.00$ & Poor \\
$0.00-0.20$ & Slight \\
$0.21-0.40$ & Fair \\
$0.41-0.60$ & Moderate \\
$0.61-0.80$ & Substantial \\
$0.81-1.00$ & Almost perfect \\
\bottomrule
\end{tabular}
\end{table}

\section{Package Design \& Core Functions}

\subsection{Overall Architecture}

The \texttt{catekappa} package consists of three main modules:
\begin{enumerate}
    \item \textbf{Design module} (functions: \texttt{calc\_sample\_size\_kappa}, \texttt{kappa\_fixed\_n}): wraps \texttt{kappaSize} functions \texttt{PowerBinary}, \texttt{Power3Cats}, \texttt{Power4Cats}, \texttt{Power5Cats} and their fixed-N counterparts.
    \item \textbf{Analysis module} (function: \texttt{analyze\_kappa}): wraps \texttt{irr} functions \texttt{kappa2}, \texttt{kappam.fleiss}, \texttt{kappam.light}.
    \item \textbf{Shiny GUI} (function: \texttt{run\_cate\_app}): launches an interactive web application built with \texttt{shiny} and \texttt{bslib}.
\end{enumerate}
The package imports \texttt{irr}, \texttt{kappaSize}, \texttt{shiny}, and \texttt{bslib}, and uses S3 classes for consistent printing (\texttt{cate\_design}, \texttt{cate\_analysis}).

\subsection{Core Export Functions}

\begin{itemize}
    \item \texttt{calc\_sample\_size\_kappa(kappa0, kappa1, props, alpha, power, raters)} \\
    Returns required sample size and parameters. \\
    \texttt{props}: numeric vector of category proportions summing to 1. \\
    \texttt{raters}: number of raters (2-6, but \texttt{kappaSize} effectively supports up to 5 categories, raters up to 6). \\
    Returns list with \texttt{n}, \texttt{kappa0}, \texttt{kappa1}, \texttt{props}, etc.

    \item \texttt{analyze\_kappa(data, type, detail)} \\
    Computes kappa statistic. \texttt{data}: data frame (rows=subjects, columns=raters). \\
    \texttt{type}: \texttt{"cohen"} (requires 2 columns), \texttt{"fleiss"} ($\ge 3$ columns), or \texttt{"light"} ($\ge 3$ columns). \\
    Returns S3 object of class \texttt{cate\_analysis} with \texttt{statistic}, \texttt{kappa}, \texttt{interpretation}, etc.

    \item \texttt{run\_cate\_app(port, launch.browser, host)} \\
    Launches the Shiny app. The app is stored in \texttt{system.file("shinyapp", package="catekappa")}.
\end{itemize}

\subsection{Shiny App Layout}

The GUI (Figure~\ref{fig:1},Figure~\ref{fig:2}) has two tabs:

\begin{itemize}
    \item \textbf{Sample Size Design}: Input fields for $\kappa_0$, $\kappa_1$, category proportions (comma-separated string), $\alpha$, power, number of raters. Clicking ``Calculate'' calls \texttt{calc\_sample\_size\_kappa} and displays the result in a color-coded box (orange if $N>200$, green otherwise) and a detailed table.

    \begin{figure}[H]
        \centering
        \includegraphics[width=0.5\linewidth]{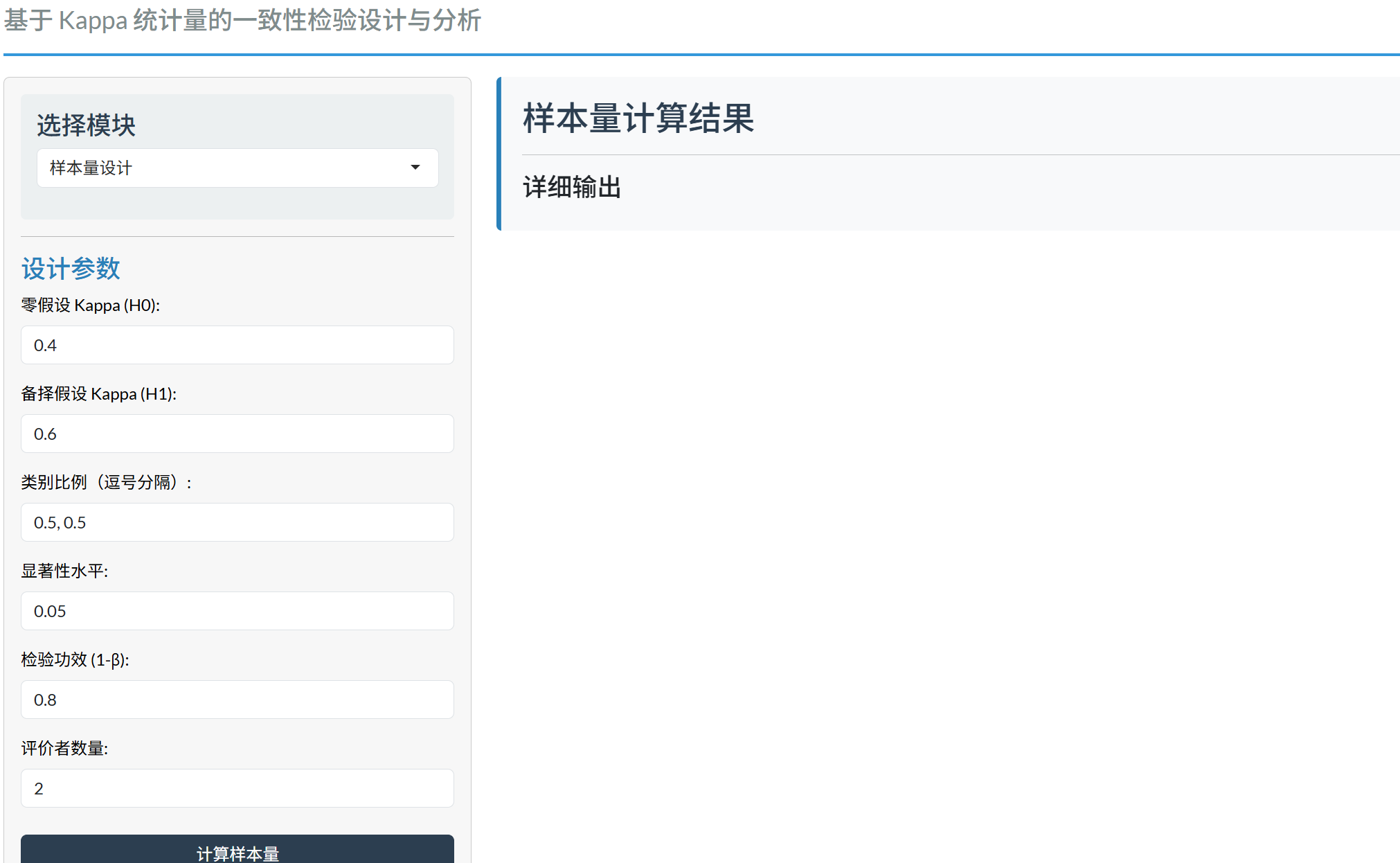}
        \caption{The CATEKAPPA Shiny application: Sample size calculation}
        \label{fig:1}
    \end{figure}
    \item \textbf{Consistency Analysis}: File upload (CSV/TXT). User chooses kappa type. Clicking ``Analyze'' calls \texttt{analyze\_kappa} and shows the estimated kappa, the Landis \& Koch level with a color indicator, and a full interpretation box containing the standard scale legend.
    
    \begin{figure}[H]
    \centering
    \includegraphics[width=0.5\linewidth]{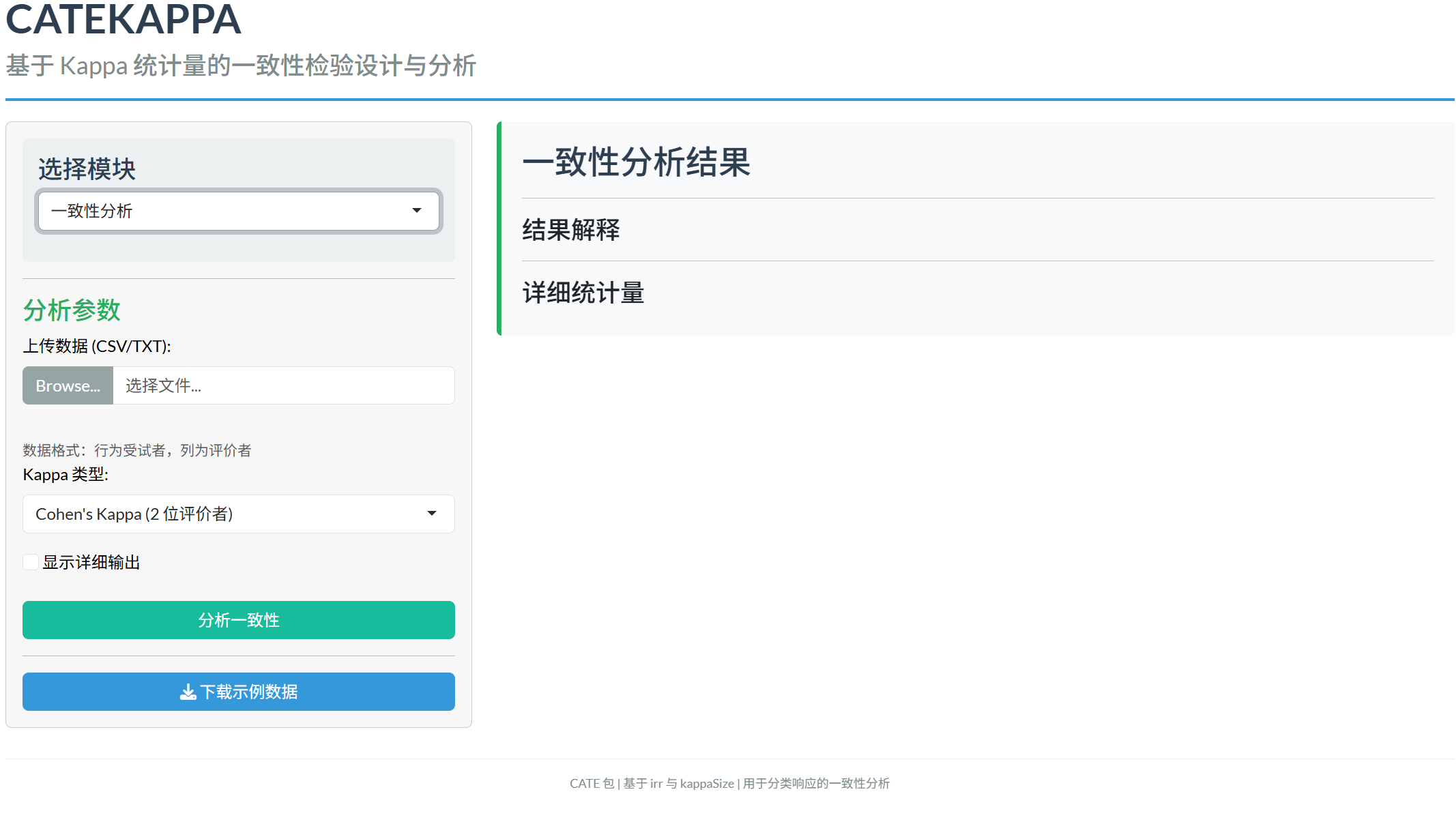}
    \caption{The CATEKAPPA Shiny application: Consistency analysis}
    \label{fig:2}
\end{figure}
\end{itemize}

\section{Usage Examples}

All examples are reproducible in an R session after installing the package.

\subsection{Sample Size Calculation}

\noindent\textbf{Example 1:} Two raters, two categories (balanced), $\kappa_0=0.4$, $\kappa_1=0.6$, $\alpha=0.05$, power=0.8.

\begin{verbatim}
library(catekappa)
calc_sample_size_kappa(kappa0 = 0.4, 
    kappa1 = 0.6,
    props = c(0.5, 0.5), alpha = 0.05,
    power = 0.8, raters = 2)
\end{verbatim}
\textbf{Output:} $n=82$ subjects (rounded up).

\noindent\textbf{Example 2:} Three raters, three categories (0.5,0.3,0.2), same $\kappa$ values, $\alpha=0.05$, power=0.8.

\begin{verbatim}
calc_sample_size_kappa(kappa0 = 0.4,
kappa1 = 0.6, props = c(0.5, 0.3, 0.2),
 alpha = 0.05, power = 0.8, raters = 3)
\end{verbatim}
\textbf{Output:} $n=99$ (requires \texttt{kappaSize::Power3Cats}).

\subsection{Agreement Analysis}

\noindent\textbf{Simulated data:} 30 subjects, 2 raters, 3 categories.

\begin{verbatim}
set.seed(123)
true <- sample(c("Low","Medium","High"), 
30, replace=TRUE,prob = c(0.3,0.4,0.3))
rater2 <- ifelse(runif(30) < 0.8, true,
        sample(c("Low","Medium","High"), 30,
        replace = TRUE))
data <- data.frame(Rater1 = true,
        Rater2 = rater2)
res <- analyze_kappa(data, 
        type = "cohen", detail = TRUE)
print(res$kappa)          
# 0.752
print(res$interpretation) 
# Substantial agreement
\end{verbatim}

\noindent\textbf{Using built-in example data (multiple sclerosis, Landis \& Koch 1977):} The package does not bundle the data by default, but users can reproduce the analysis from the original paper. For demonstration, we create a small synthetic dataset:

\begin{verbatim}
# 4 categories, 2 raters, 149 subjects
set.seed(456)
cat4 <- c("Certain",
"Probable","Possible","Doubtful")
true4 <- sample(cat4, 149, 
    replace = TRUE,
    prob = c(0.56,0.25,0.07,0.12))
rater2_4 <- ifelse(runif(149) < 0.7, true4,
sample(cat4, 149, replace = TRUE))
data4 <- data.frame(Rater1 = true4,
    Rater2 = rater2_4)
analyze_kappa(data4, type = "cohen")
\end{verbatim}
\begin{figure}[H]
    \centering
    \includegraphics[width=0.5\linewidth]{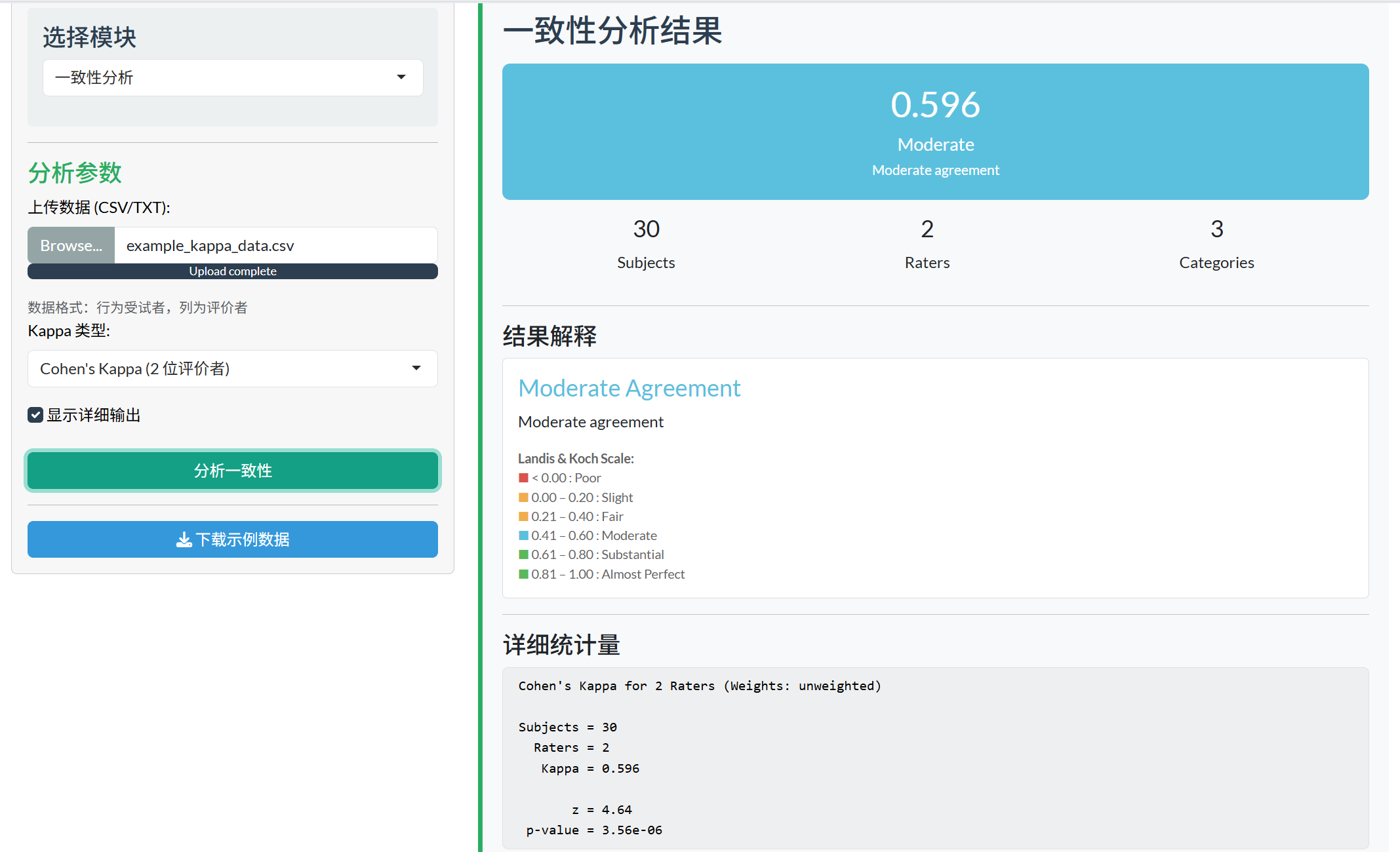}
    \caption{Consistency analysis result}
    \label{fig:3}
\end{figure}
\begin{figure}[H]
    \centering
    \includegraphics[width=0.5\linewidth]{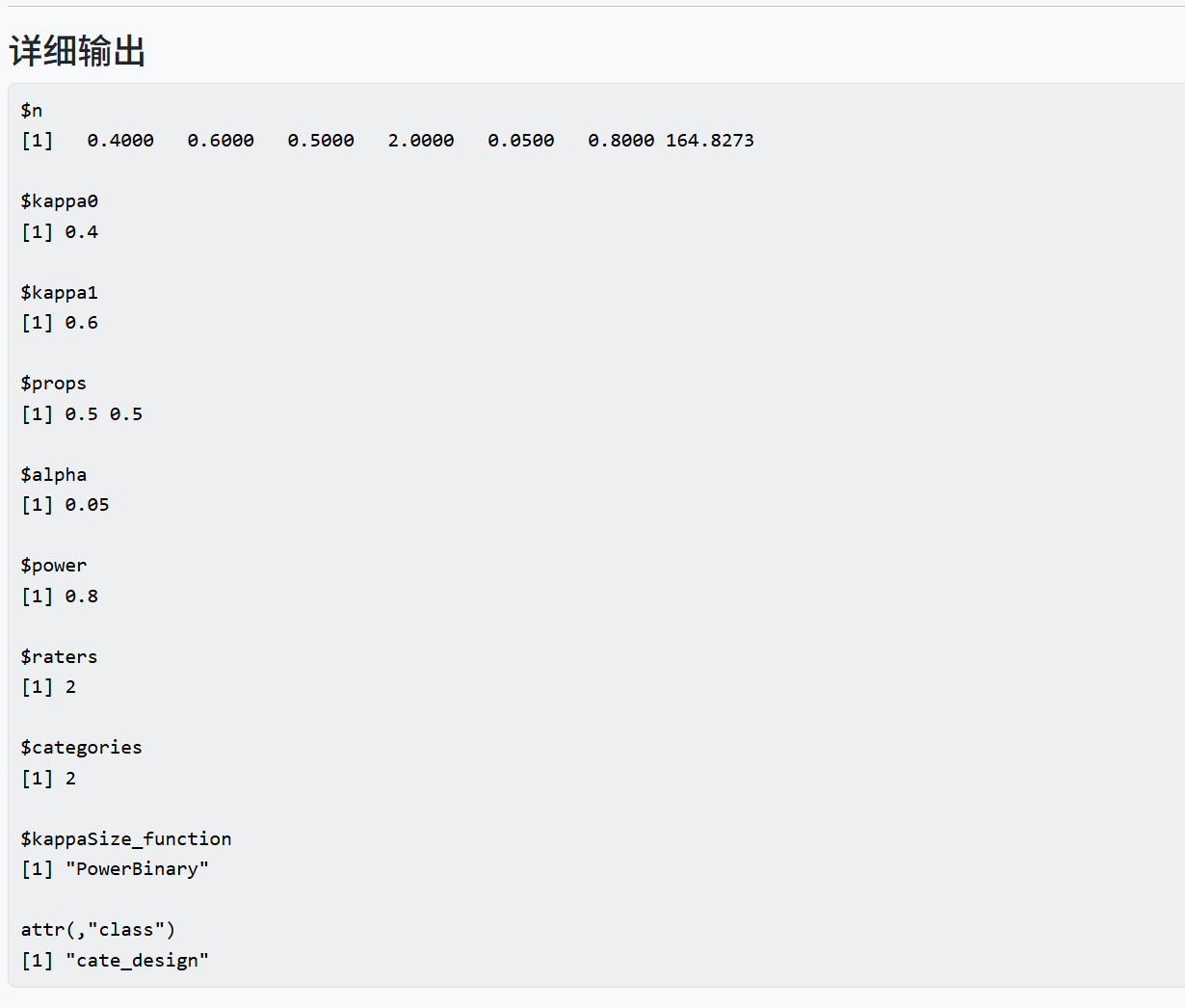}
    \caption{Sample size calculation result}
    \label{fig:4}
\end{figure}
\subsection{Launching the Shiny App}

\begin{verbatim}
run_cate_app()
\end{verbatim}
The app will open in the default web browser. Figure~\ref{fig:3},Figure~\ref{fig:4}  show a typical analysis result.

\section{Discussion}

\texttt{CATEKAPPA} provides an integrated, user-friendly solution for two common tasks in categorical agreement studies: sample size planning and post-hoc kappa analysis. By wrapping established R packages (\texttt{irr}, \texttt{kappaSize}) and providing a Shiny interface, it lowers the barrier for non-programmers while retaining full functionality for advanced users via command-line functions.

\paragraph{Limitations.}
The sample size calculations rely on large-sample approximations, which may be inaccurate for small $N$ (e.g., $N<50$). Users should treat results as approximate when the expected sample size is small. Additionally, the package currently supports only unweighted kappa; weighted kappa for ordinal categories is not implemented. The \texttt{kappaSize} package also restricts category numbers to 2–5; for more than 5 categories, the sample size function will throw an error.

\paragraph{Future work.}
We plan to extend the package to:
\begin{itemize}
    \item Include weighted kappa and associated sample size formulas.
    \item Add Bayesian kappa estimation with credible intervals.
    \item Support for more than 5 categories using alternative methods (e.g., bootstrap-based sample size determination).
    \item Provide additional plotting functions (e.g., agreement heatmaps, confidence interval plots).
\end{itemize}

\section{Conclusion}

We have developed \texttt{CATEKAPPA}, an R package with an accompanying Shiny application that facilitates sample size determination and agreement analysis based on the kappa statistic for categorical responses. The package is freely available on CRAN and provides both a graphical interface and command-line tools. We believe it will serve as a valuable resource for researchers in medicine, psychology, and social sciences who need to design and analyze inter-rater reliability studies.

\section*{Acknowledgements}

The authors thank the maintainers of the \texttt{irr}, \texttt{kappaSize}, and \texttt{shiny} packages for their foundational work. We are also grateful to the CRAN team for their review and acceptance of the package. No specific funding was received for this work.

The package source code is available at \url{https://github.com/satellite837/catekappa}. The CRAN release page is \url{https://CRAN.R-project.org/package=catekappa}.

\bibliographystyle{plainnat}

\appendix
\section{Appendix: Full Package Function List}

\noindent\textbf{Exported functions:}
\begin{itemize}
    \item \texttt{calc\_sample\_size\_kappa()}: sample size calculation.
    \item \texttt{kappa\_fixed\_n()}: given sample size, compute lower confidence bound.
    \item \texttt{analyze\_kappa()}: compute kappa and interpretation.
    \item \texttt{run\_cate\_app()}: launch Shiny app.
    \item \texttt{interpret\_kappa()}: internal but exported; returns Landis \& Koch classification.
\end{itemize}

\noindent\textbf{Internal (non-exported) helper:}
\begin{itemize}
    \item \texttt{check\_kappa\_data()}: validates input data.
\end{itemize}

\noindent\textbf{S3 classes:}
\begin{itemize}
    \item \texttt{cate\_design} (returned by \texttt{calc\_sample\_size\_kappa}) with \texttt{print} method.
    \item \texttt{cate\_analysis} (returned by \texttt{analyze\_kappa}) with \texttt{print} and \texttt{summary} methods.
    \item \texttt{cate\_fixed\_n} (returned by \texttt{kappa\_fixed\_n}) with \texttt{print} method.
\end{itemize}

\noindent\textbf{Shiny app structure:}
The app is located in \texttt{inst/shinyapp/} and consists of a single \texttt{app.R} file (combined UI and server). It uses \texttt{bslib} for theming. Inputs are passed to \texttt{calc\_sample\_size\_kappa} or \texttt{analyze\_kappa}, and outputs are rendered via \texttt{renderUI} and \texttt{verbatimTextOutput}.

\noindent\textbf{Minimum supported R version:} 3.5.0.

\noindent\textbf{The demo code is as follows:}
\begin{verbatim}
# Load the package
library(catekappa)

# Sample size calculation (2 raters, 2 categories)
calc_sample_size_kappa(
  kappa0 = 0.4, kappa1 = 0.6,
  props = c(0.5, 0.5), alpha = 0.05,
  power = 0.8, raters = 2
)

# Sample size calculation (3 raters, 3 categories)
calc_sample_size_kappa(
  kappa0 = 0.4, kappa1 = 0.6,
  props = c(0.5, 0.3, 0.2),
  alpha = 0.05, power = 0.8, raters = 3
)

# Simulate agreement data (2 raters, 3 categories)
set.seed(123)
true_class <- sample(c("Low", "Medium", "High"),
    30,replace = TRUE, prob = c(0.3, 0.4,0.3))
rater2 <- ifelse(runif(30) < 0.8, true_class,
        sample(c("Low", "Medium", "High"), 30,
        replace = TRUE))
agreement_data <- data.frame(Rater1 = true_class,
                             Rater2 = rater2)

# Compute Cohen's kappa
result <- analyze_kappa(agreement_data, type = "cohen")
print(result$kappa)
print(result$interpretation$level)
\end{verbatim}

\end{document}